\documentclass{jpsj3}
\usepackage{txfonts}
\usepackage{bm}
\usepackage{xcolor}

\DeclareUnicodeCharacter{0301}{\'{}}

\title{Bayesian Inference for Small-Angle Scattering Data II: Core–Shell Samples}

\author{Keigo Oyama$^1$, Yui Hayashi$^1$, Shigeo Kuwamoto$^2$, Shun Katakami$^1$, Kenji Nagata$^3$, Masaichiro Mizumaki$^4$, Masato Okada$^1$\thanks{okada@edu.k.u-tokyo.ac.jp}}
\inst{$^1$Graduate of Frontier Sciences, The University of Tokyo, Kashiwa, Chiba 277-8561, Japan \\
$^2$ Japan Synchrotron Radiation Research Institute (JASRI), Sayo, Hyogo 679-5198, Japan \\
$^3$ Center for Basic Research on Materials (CBRM), National Institute for Materials Science (NIMS), Tsukuba, Ibaraki 305-0044, Japan \\
$^4$ Faculty of Advanced Science and Technology, Kumamoto University, Kumamoto, Kumamoto 860-8555, Japan} 

\abst{Small-angle scattering (SAS) experiments, which utilize X-rays or neutron beams, are employed in various scientific fields, including materials science, biochemistry, and polymer physics. During the analysis of SAS data, model parameters that contain information about the sample are estimated by fitting the measurement data to a model of sample. Previous research has demonstrated the effectiveness of Bayesian inference in analyzing SAS data using a sphere model. However, compared with the sphere model, the core-shell model, which represents functional nanoparticles, offers higher application potential and greater analytical value. Therefore, in this study, we propose an analytical method for the more complex and practical core-shell model based on Bayesian inference. Through numerical experiments, we evaluated the performance of this method under different conditions, including measurement times, number of data points, and differences in scattering length density. As a result, we clarify the conditions under which accurate estimations are possible.}

\begin{document}
\maketitle
    
\section{Introduction}
    Small-angle scattering (SAS) is a experiment that analyzes materials by exposing them to X-rays or neutron beams and measuring the scattered radiant intensity as a function of the scattering angle, typically within a range below 5 degrees \cite{guinier1979small}. This technique allows for the estimation of structural parameters, such as size and density, by fitting a model that represents the scattering intensity to the collected scattering data.
    Traditional fitting methods, such as nonlinear least squares face challenges, including the risk of converging to local minima, difficulties in objectively assessing the confidence of the estimates, and sensitivity to initial values, which can affect the accuracy of the results.

    Previous studies have addressed these problems by applying Bayesian inference for parameter estimation to synthetic data of both monodisperse and polydisperse spheres \cite{hayashi2023bayesian}. The Bayesian approach allows for the evaluation of parameter confidence and the avoidance of local minima by employing the Replica-Exchange Monte Carlo(EMC) method \cite{nagata2012bayesian}. This method effectively addresses the challenges of traditional techniques, such as least squares or gradient-based optimization methods \cite{szymusiak2017core}\cite{pedersen1997analysis} \cite{shi2023small} , by evaluating parameter confidence and avoiding local minima. However, these studies have largely been limited to theoretical descussions with simple spherical model. We focused on core-shell samples, which are of significant practical importance and have many applications, including their use in lithium-ion batteries \cite{zhu2020situ} and drug delivery systems \cite{krycka2013resolving}.
    Core-shell samples introduce a complex optimization challenge due to the numerous free parameters involved and the inclusive nature of the core-shell model, which includes the sphere model. In practical scenarios, theoretical calculations and assumptions are often used to fix certain parameters as constants, thereby simplifying the analysis by reducing the number of parameters considered \cite{tian2020structural}.

    We explored the limits of our parameter estimation approach for core-shell models through numerical expriments under three different scenarios. We adjusted the parameters for the inner radius and outer radius, scattering length densities of the core/shell, and background using artificial data. In this paper, the particle size distribution is not considered in order to keep the discussion simple. However, if one wishes to take particle size distribution into account, it can be modeled by convolving the typical size distribution with a shape factor. \cite{hayashi2023bayesian}\cite{sivia2011elementary}

    In the first scenario, we investigated how varying measurement duration impacts the estimation accuracy. Given the trade-off between the number of data points and associated costs in SAS, our aim was to assess the limits of Bayesian inference concerning the quantity of data collected. The second and third conditions analyzed the effects of varying the measurement range of the scattering vector $\bm{q}$. We systematically reduced the measurement range from both the wide-angle and low-angle sides to evaluate the estimation limits in each case.

    The structure of this paper is outlined as follows: Sect. \ref{sec:formulation} presents theoretical framework for the core--shell model used in the numerical experiments. Sect. \ref{sec:framework} describes the proposed method. Sect. \ref{sec:numerical_experiment} provides an overview of the numerical experiments, and their results. Finally, Sect. \ref{sec:conclusion} presents the conclusions drawn from the study.

\section{Formulation of Core--Shell Models}

    \label{sec:formulation}
\begin{figure}[htbp]
            \centering
            \includegraphics[keepaspectratio, scale=0.7]{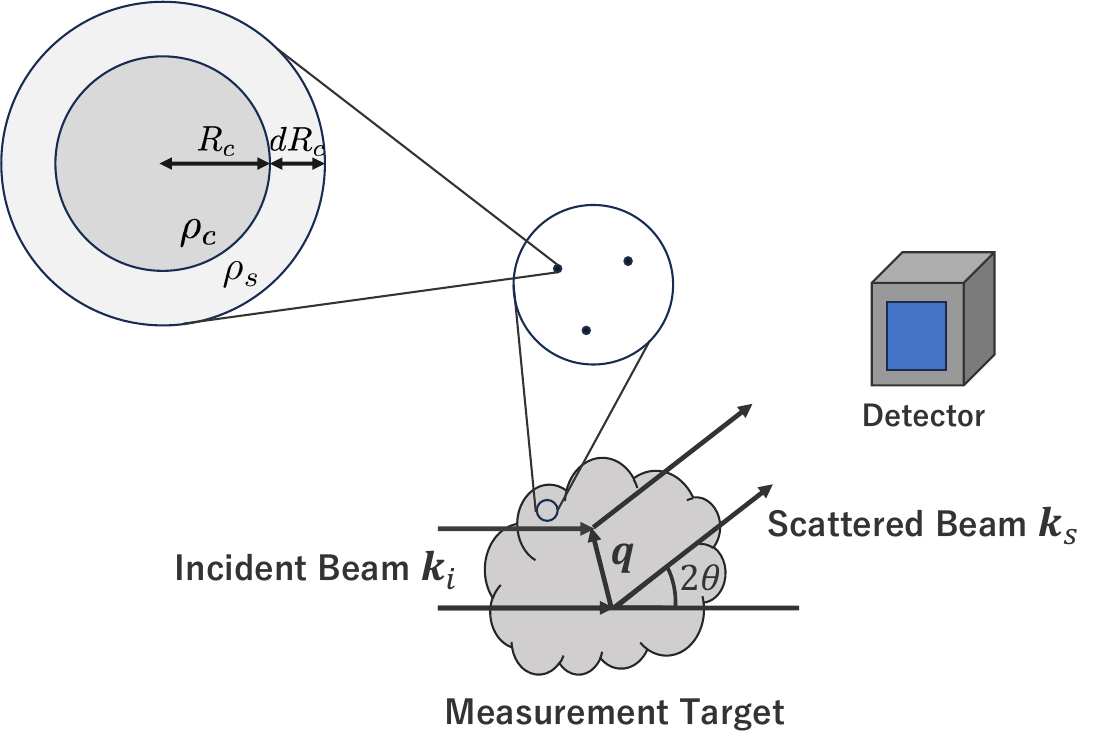}
            \caption{\normalsize Schematic diagram of SAS}
        \label{fig:explain_sas}
    \end{figure}

In this section, we outline the process for generating SAS data. We consider a core-shell model to represent the sample. Then, we derive the scattering intensity and its measurements for this model.
    A core-shell particle consists of a two-layer structure: the core is a sphere of radius $R_c$ with a scattering length density $\rho_c$, surrounded by an outer shell of thickness $dR_c$ and scattering length density $\rho_s$, as shown in Fig. \ref{fig:explain_sas}.

    First, we derive the scattering intensity for core-shell particles. Let $\bm{q}$ represent the scattering vector of the scattered wave, $\bm{k}_i$ be the wavenumber vector of the incident light, $\bm{k}_s$ be the wavenumber vector of the scattered light, and $\lambda$ be the wavelength of the incident wave. Then, the angle between $\bm{k}_i,\bm{k}_s$ is $2\theta$ and $\bm{q}$ is defined by Eq. (\ref{eq:define_q}).
    \begin{align}
        \bm{q} \equiv \frac{4 \pi \sin \theta\left(\bm{k}_s-\bm{k}_i\right)}{\lambda | \bm{k}_s-\bm{k}_i |}.
        \label{eq:define_q}
    \end{align}
    Assuming isotropic scattering and considering only the magnitude of the scattering vector, the scattering intensity $I(q;\Theta, t)$ for core-shell particles can be expressed as in Eq. (\ref{eq:define_I}).\cite{pedersen1997analysis}
    \begin{align}
        I(q;\Theta, t)&= t*\left(\frac{A}{V_s}\left[\frac{3 V_c\left(\rho_c-\rho_s\right) j_1\left(q R_c\right)}{q R_c}+\frac{3 V_s\left(\rho_s-\rho_{solv}\right) j_1\left(q R_s\right)}{q R_s}\right]^2+b \right) ,
        \label{eq:define_I}
    \end{align}
    \begin{align}
        R_s&=R_c+d R_c, \\
        V_s&=(4 \pi / 3)\left[R_s^3-R_c^3\right], \\
        V_c&=(4 \pi / 3) R_c^3, \\
        j_1(x)&=\frac{(\sin x-x \cos x)}{x^2}.
    \end{align}
    where $\Theta$ represents the parameters of core-shell samples, $t$ is the measurement time in seconds, $b$ is the background noise, and $A$ is the scale factor. Assuming the sample is sufficiently dilute, we approximate the structure factor as 1.
    Our measurement of $y$ is the scattered intensity $I(q;\Theta, t)$ with some noise. As the scattering intensity corresponds to the number of photons detected by the detector, we model Poisson noise as the noise. Therefore, the measured scattered intensity $y$ can be expressed using Eq. (\ref{eq:define_y}).\cite{nagata2019bayesian}
    \begin{align}
        y \overset{i.i.d}{\sim} p(y) =\frac{e^{-I(q;\Theta, t)} I(q;\Theta, t)^y}{y !} .
        \label{eq:define_y}
    \end{align}

\section{Bayesian Framework}
\label{sec:framework}
    This section describes the Bayesian framework for estimating parameters in SAS, as described by Hayashi et al.\cite{hayashi2023bayesian}.

\subsection{Bayesian Inference for Parameters in SAS}
    This subsection describes the Bayesian inference method employed for parameter estimation. Bayes’ theorem, as shown in Eq. (\ref{eq:define_bayesian}), aims to determine the posterior $p(\Theta \mid D)$ of the parameters.

    \begin{align}
        p(\Theta \mid D) &=\frac{p(D \mid \Theta)p(\Theta)}{p(D)},
        \label{eq:define_bayesian}
    \end{align}

    The measured data are $D=\{q_i, y_i\}_{i=1}^N$ and the parameters of the core-shell model sample are $\Theta$. The likelihood $p(y \mid \Theta)$ is expressed as Eq. (\ref{eq:define_y}), while the prior distribution $p(\Theta)$ is determined based on prior knowledge. Since $p(D)$ cannot be calculated analytically, the posterior obtained by sampling from the right-hand side of Eq. (\ref{eq:define_proportional_posterior}).
    \begin{align}
        p(\Theta \mid D) \propto p(D \mid \Theta)p(\Theta).
        \label{eq:define_proportional_posterior}
    \end{align}
    The right-hand side of Eq. (\ref{eq:define_proportional_posterior}) can be reformulated as Eq. (\ref{eq:change_likelihood}), incorporating a cost function such as Eq. (\ref{eq:define_cost_function}), where the measured scattering intensity $y$ is assumed to be independently generated based on the parameter $\Theta$. Subsequently, Poisson noise is applied as the noise.
    \begin{align}
        p(D \mid \Theta)p(\Theta) & =\prod_{i=1}^N p\left(y_i \mid q_i, \Theta\right)p(\Theta), \\
        & =\exp (-N E(\Theta))p(\Theta), \label{eq:change_likelihood} \\
        E(\Theta) &\equiv \frac{1}{N} \sum_{i=1}^N\left\{I(q_i ; \Theta)-y_i \log I(q_i ; \Theta)+\sum_{j=1}^{y_i} \log j\right\}. \label{eq:define_cost_function}
    \end{align}

\subsection{EMC Method}
    The EMC method was used as the sampling instrument.
    The sampling was performed according to the following method.
    \begin{enumerate}
        \item Prepare $M$ temperature layers and determine the initial value $\Theta_m\, (m=1,\dots,M)$ of each temperature layer. The inverse temperature $\beta_m\, (m=1,\dots,M)$ is defined as $\beta_0 = 0 < \beta_1 < \dots < \beta_M = 1$ and the likelihood and posterior at temperature layer $m$ is defined by Eq. (\ref{eq:define_emc_proportional_posterior}).
        \begin{align}
            p_{\beta_m}(\Theta_m \mid D) \propto \exp(- \beta_m N E(\Theta_m))p(\Theta_m).
            \label{eq:define_emc_proportional_posterior}
        \end{align}
        \item Update $\Theta_m$ for each temperature layer by MCMC method using the Metropolis method \cite{bhanot1988metropolis}.
        \item In the temperature layer $\beta_m, \beta_{m+1}$, calculate Eq. (\ref{eq:define_exchange_rate}).
        \begin{align}
            & \alpha\left(\Theta_{m}, \Theta_{m+1}\right)=\min (1, v) \label{eq:define_exchange_rate}, \\
            & v=\frac{p_{\beta_m}\left(\Theta_{m+1} \mid D\right) p_{\beta_{m+1}}\left(\Theta_m \mid D\right)}{p_{\beta_m}\left(\Theta_m \mid D\right) p_{\beta_{m+1}}\left(\Theta_{m+1} \mid D\right)}, \\
            & =\exp \left(N\left(\beta_{m+1}-\beta_m\right)\left(E\left(\Theta_{m+1}\right)-E\left(\Theta_m\right)\right)\right).
        \end{align}
        \item Starting from the highest temperature layer, the parameter values $\Theta_m$ and $\Theta_{m+1}$ of the adjacent temperature layers $m$ and $m+1$ are swapped with a probability $alpha$.
        \item $(2) \sim (4)$ repeat.
    \end{enumerate}

\subsection{Maximum a Posteriori}
    Bayesian inference enables the estimation of parameter values and intervals by sampling from the posterior. The maximum a posteriori (MAP) estimation, denoted as $\Theta_{\rm{MAP}}$ in Eq. (\ref{eq:define_map}), provides a point estimate of the parameters, representing the combination with the highest posterior.
    \begin{align}
        \Theta_{\rm{MAP}} = \underset{\Theta_M} {\operatorname{argmax}} \,  p_{\beta_M} (\Theta_M |D)   
        \label{eq:define_map}
    \end{align}

\section{Numerical Experiment}
\label{sec:numerical_experiment}
    In this section, we present the procedure and results of our numerical experiments conducted using synthetic data. These experiments involved estimating model parameters from synthetic data generated using the core-shell model described in Sec. \ref{sec:formulation} and applying the method presented in Sec. \ref{sec:framework}.

    We begin by outlining the procedure for generating synthetic data. First, $N$ magnitudes $q$ of the scattering vectors are given at equal intervals in the interval $[q_{\rm{min}}, q_{\rm{max}}]$. For each $q$ value, we calculate the measured scattering intensity $y$ using Eq. (\ref{eq:define_y}). The synthetic data $D=\{{q_i, y_i}\}^N_{i=1}$ are obtained from the above procedure. We set $q_{\rm{min}}=0.001\,\rm{nm}^{-1}$,\,$q_{\rm{max}}=0.7\,\rm{nm}^{-1}$,\,$N=500$. The data are analyzed for the measurement times $t=1000$, except for Subsec. \ref{subsec:condition1}.

    The model parameters for the core-shell samples are described. Assuming that the model parameters for a core-shell sample are the core radius $R_c$, shell thickness $dR_c$, core scattering length density $\rho_c$, shell scattering length density $\rho_s$, and background $b$, $\Theta = \left\{R_c\,({\rm{nm}}), dR_c\,({\rm{nm}}), \rho_c\,(\times 10^{-4}\,{\rm{nm}^{-2}}), \rho_s\,(\times 10^{-4}\,{\rm{nm}^{-2}}), b\,(\rm{cm}^{-1})\right\}$. The solvent scattering length density $\rho_{solv}$ is $\rho_{solv}=0$ with a scale factor $A=1$. The true values of the model parameters $\Theta^*$ are expressed as $\Theta^* = \left\{R_c^*, dR_c^*, \rho_c^*, \rho_s^*, b^*\right\} = \{60, 10, 0.001, 1.0, 1.2\}$. 
    The prior $p(\Theta)$ for the model parameter $\Theta$ was empirically determined. The initial values of $\Theta$ were obtained as random numbers according to the prior assigned to each parameter.
    \begin{align}
        p(\Theta) &= p(R_c) p(dR_c) p(\rho_c) p(\rho_s) p(b), \\
        p\left(R_c\right) & =\operatorname{Gamma}\left(R_c ; \alpha_R, \gamma_R\right), \\
        & =\frac{e^{-x / \gamma_{R}}}{\gamma_R^{\alpha_R} \cdot \Gamma\left(\alpha_R\right)} \cdot x^{\alpha_R-1}, \\
        p(dR) & =\operatorname{Gamma}\left(dR ; \alpha_{dR}, \gamma_{dR}\right), \\
        p(b) & =\operatorname{Gamma}\left(b ; \alpha_b, \gamma_b\right) , \\
        p(\rho_c) & =\operatorname{Gamma}\left(\rho_c ; \alpha_{\rho_c}, \gamma_{\rho_c}\right), \\
        p(\rho_s) & =\operatorname{Gamma}\left(\rho_s ; \alpha_{\rho_s}, \gamma_{\rho_s}\right),
    \end{align}

    \begin{align}
        \alpha_R & =20, \gamma_R=3, \\
        \alpha_{dR} & =1.05, \gamma_{dR}=100, \\
        \alpha_b & =1.05, \gamma_b=5, \\
        \alpha_{\rho_c} & =1.05, \gamma_{\rho_c}=30, \\
        \alpha_{\rho_c} & =1.05, \gamma_{\rho_s}=30.
    \end{align}

    The number of EMC replicas was fixed at $M=70$, with the inverse temperature $\beta_m(m=1,2,\dots,M)$ for each replica being assigned according to a specified method. The results presented in this section were derived with an EMC burn-in period of $10^5$ and a subsequent sampling period of $10^5$.
    \begin{align}
        \beta_m =
        \left\{
            \begin{array}{ll}
                0 & (m=1) \\
                1.5^{m-M} & (m \geq 2)
            \end{array}
        \right. .
    \end{align}

    The parameter estimation was compared under three specific conditions. The conditions were as follows.

    \subsection{Assessing Estimation Accuracy by Varying Measurement Times}    
    \label{subsec:condition1}
        In this subsection, the measurement times $t$ was varied.

\begin{figure}[htbp]
            \centering
            \includegraphics[keepaspectratio, scale=0.45]{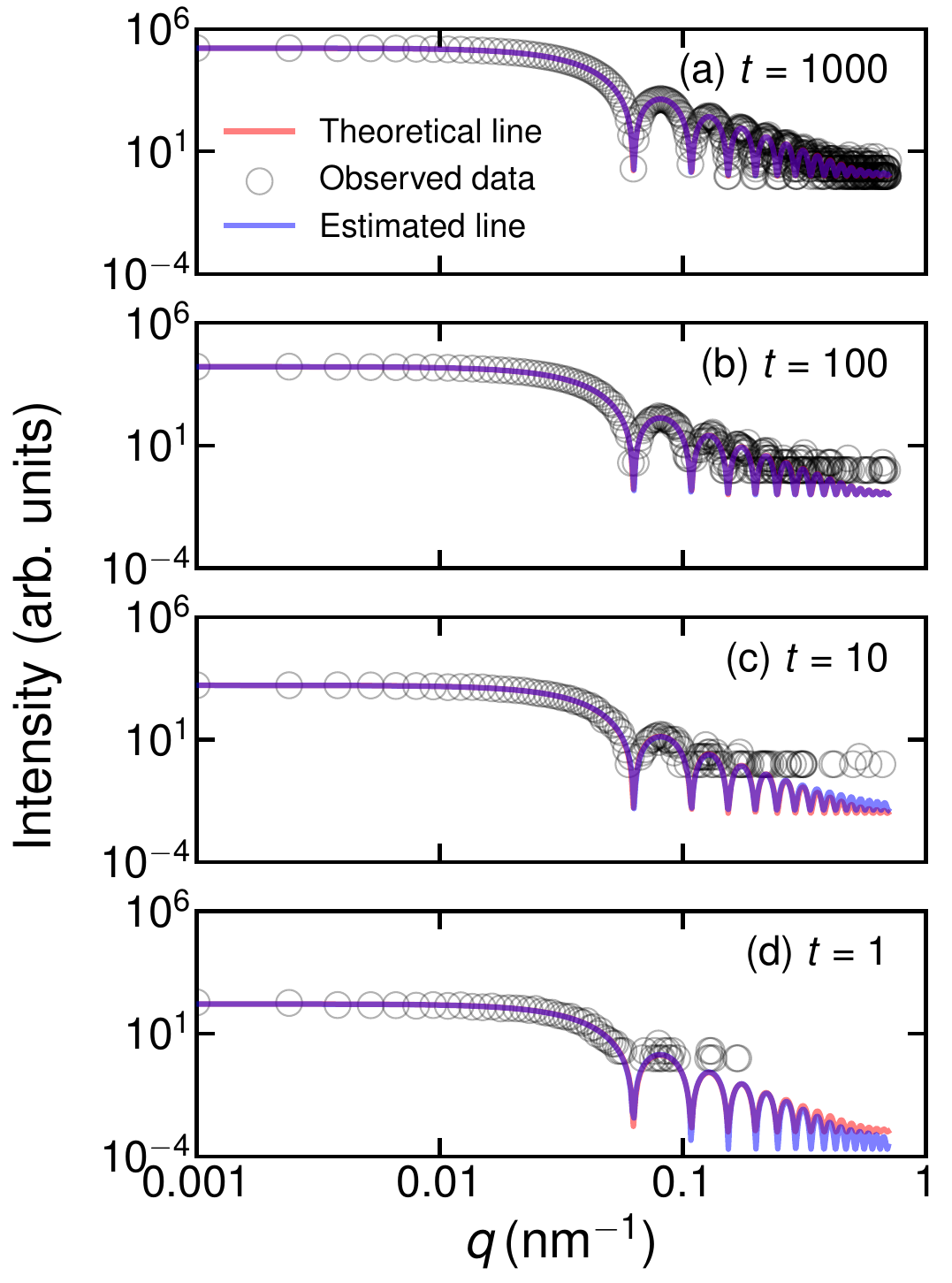}            
            \caption{Plot of synthetic data at varying measurement times $t$ and results of fitting with MAP solution. The blue line indicates the estimated curve, the red line indicates the theoretical curve, and the open black circles indicate the observed data. From top to bottom: $t=1000, 100, 10,$ and $1$.}
            \label{fig:tl_fitting_result}
        \end{figure}

\begin{figure}[htbp]
            \includegraphics[keepaspectratio, scale=0.47]{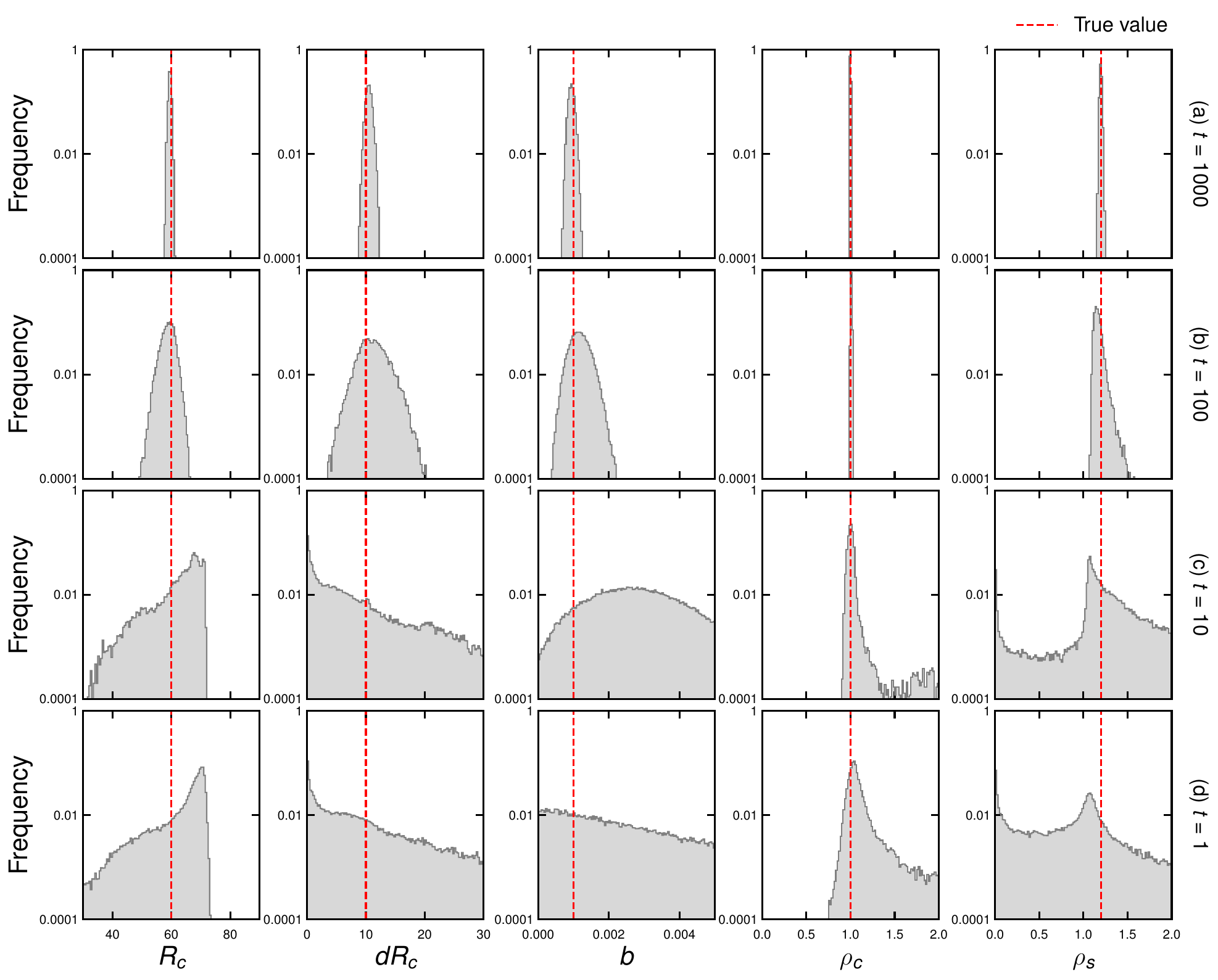}  
            \caption{Parameters $R_c, dR_c, b, \rho_c$, and $\rho_s$ are arranged from left to right, with $N=500$ samples. The posteriors at measurement times $t=1000, 100, 10,$ and $1$ are shown from top to bottom, corresponding to the labels on the right. The dashed line indicates the true values. The vertical axis is logarithmic, and 120 bins are used for the histogram.}
            \label{fig:tl_posterior_result}
        \end{figure}

        As the measurement times $t$ increased, more data were collected; however, this improvement in data quantity resulted in higher costs. Therefore, we investigated the limits of estimation by changing the measurement times. This subsection details the setup and outcomes of the numerical experiments where the measurement times $t$ were varied.
        Parameter estimation was performed on four sets of synthetic data, each generated with $q_{{\rm{min}}}=0.001\,\rm{nm}^{-1}$, $q_{\rm{max}}=0.7\,\rm{nm}^{-1}$, $N=500$ and the measurement times $t=1000,100,10,$ and $1$ using the method proposed in this paper. 
        Figure \ref{fig:tl_fitting_result} shows the plots of the synthetic data and the results of fitting the data using the MAP solution.
        Figure \ref{fig:tl_posterior_result} is the posterior of the sampled parameters.
        Figure \ref{fig:tl_fitting_result} shows that shorter measurement times result in fewer measurements.
        In Fig. \ref{fig:tl_posterior_result}, the posteriors at $t=10$ and $1$ of the parameters $R_c$ and $dR_c$ are approximately 70 and 0, respectively, whereas $\rho_c$ and $\rho_s$ peak at approximately 1.1. This suggests that the sample can be estimated as a spherical model with a radius equals 70, indicating an estimation limit between $t=10$ and $100$. For all parameters, increasing $t$ results in reduced variance in the parameter posterior, reflecting a smaller confidence interval.
        Table \ref{tab:tl_estimated_values} shows the parameter estimation results obtained from the numerical experiments in this section.

        \begin{table}[hbtp]
            \caption{Parameter estimation results. MAP solution of the posterior of each estimated model parameter $\Theta_{\rm{MAP}}$ and ${\Theta_{\rm{MAP}}}^{+p}_{-m}$, where $p,m$ is the difference between the upper and lower limits of the 99$\%$ confidence interval.}
            \label{tab:tl_estimated_values}
            \begin{tabular}{|l|c|c|c|c|c|}
                \hline
                 & $R_c$ & $dR_c$ & $ b\,(\times 10^{-3})$ & $\rho_c\,(\times 10^{-2})$ & $\rho_s\,(\times 10^{-2})$ \\
                \hline
                \hline
                $t=1000$     & $ 59.48^{+0.83}_{-0.89} $      & $ 10.50^{+0.96}_{-0.90} $    & $ 0.94^{+0.16}_{-0.14} $       & $ 99.90^{+0.29}_{-0.30} $        & $ 119.48^{+2.31}_{-1.93} $ \\
                $t=100$      & $ 59.03^{+3.74}_{-4.66} $      & $ 11.10^{+4.92}_{-4.06} $    & $ 1.16^{+0.55}_{-0.45} $      & $ 100.92^{+0.99}_{-1.00} $       & $ 115.77^{+11.16}_{-5.20} $ \\
                $t=10$      & $ 61.23^{+9.94}_{-16.78} $      & $ 9.00^{+27.66}_{-8.96} $    & $ 2.48^{+3.04}_{-1.85} $     & $ 100.38^{+82.78}_{-5.54} $   & $ 118.64^{+3952.26}_{-117.55} $ \\
                $t=1$      & $ 58.53^{+13.12}_{-16.99} $   & $ 12.47^{+136.58}_{-12.43} $   & $ 0.16^{+10.15}_{-0.07} $   & $ 103.24^{+471.44}_{-10.35} $   & $ 108.25^{+3600.04}_{-107.87} $ \\
                \hline
            \end{tabular}
        \end{table}

\newpage
\subsection{Assessing Estimation Accuracy with Reduced Data from the Wide-Angle Side}
\begin{figure}[htbp]
        \centering
        \includegraphics[keepaspectratio, scale=0.5]{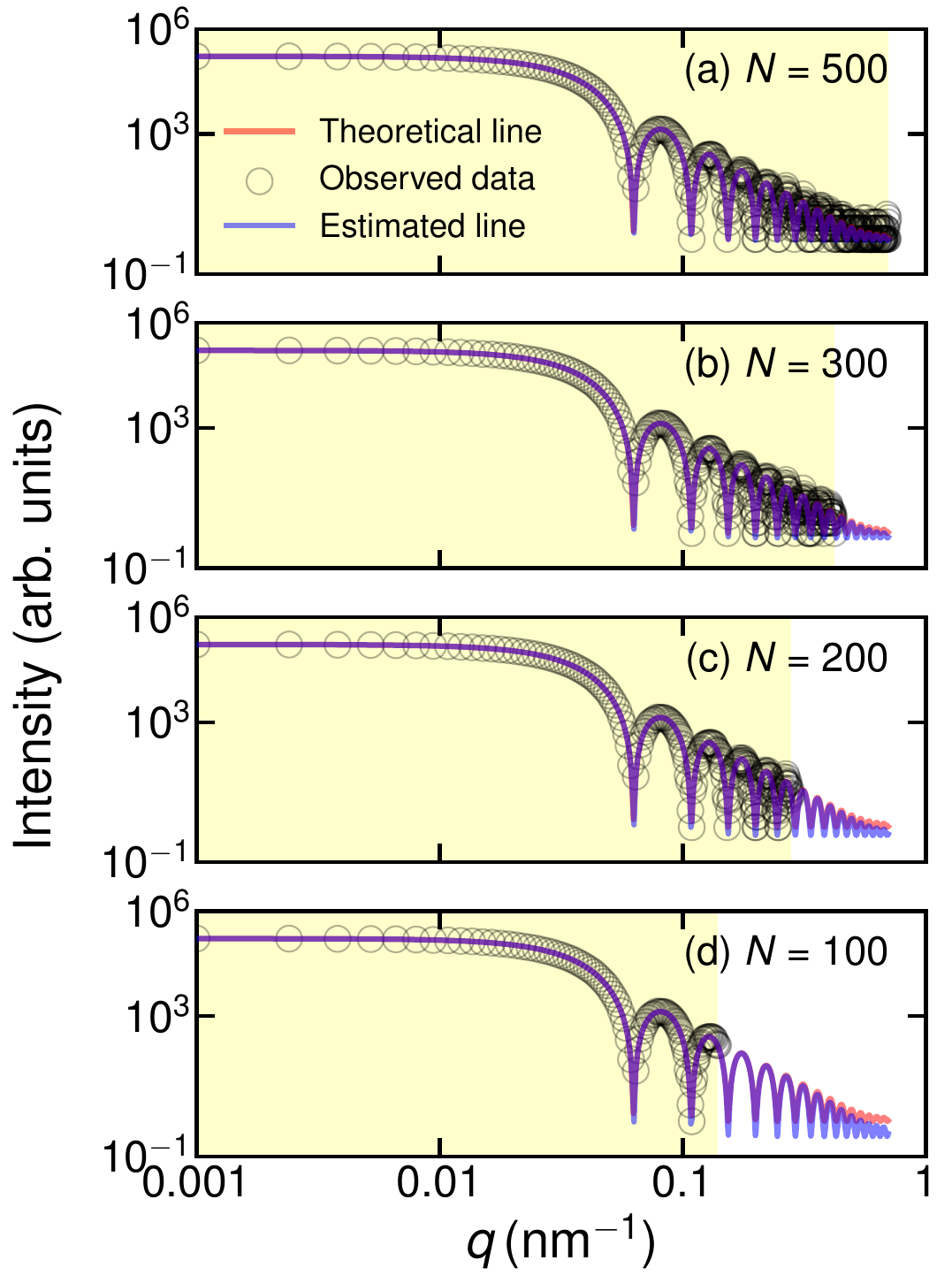}        
        \caption{Plot of synthetic data at varying values of $q_{\rm{max}}$ and results of fitting with the MAP solution. The blue line indicates the estimated curve, the red line indicates the theoretical curve, and the open black circles denote the observed data. From top to bottom: $N=500, 300, 200,$ and $100$. The yellow-shaded area indicates the range over which data is available.}
        \label{fig:hc_fitting_result}
    \end{figure}

\begin{figure}[htbp]
        \includegraphics[keepaspectratio, scale=0.47]{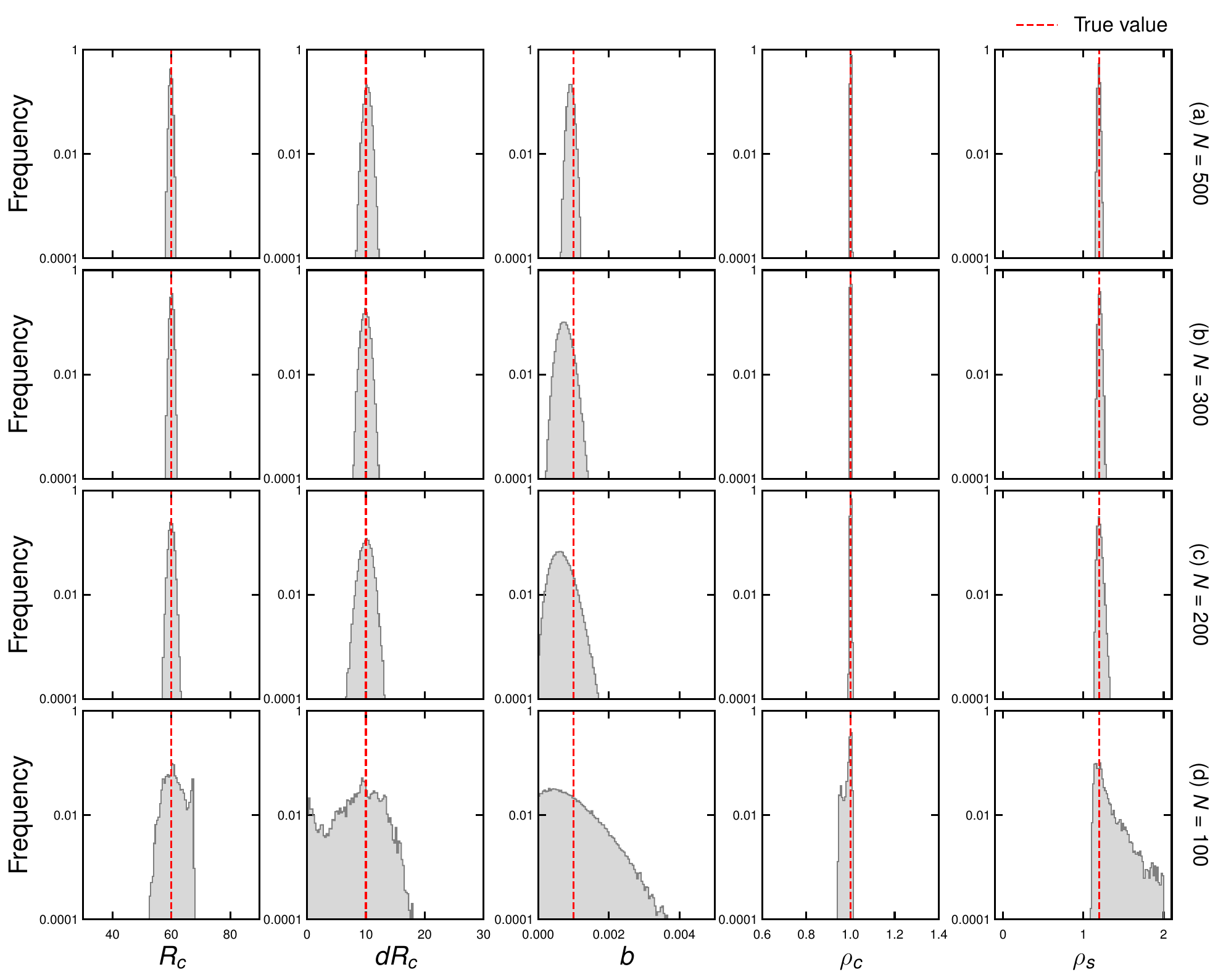}        
        \caption{Parameters $R_c, dR_c, b, \rho_c, \rho_s$ from left to right, posteriors at $N=500, 300, 200,$ and $100$ from top to bottom like right labels, and $t=1000$. The resolution is maintained constant while $q_{\rm max}$ is varied. The dashed line represents the true value, and the vertical axis is logarithmic with 120 bins used for the histogram.}
        \label{fig:hc_posterior_result}
    \end{figure}

    In this subsection, we describe the setup and outcomes of a numerical experiment where the data from the wide-angle side are shaved off.
    Initially, we removed data from the wide-angle side and conducted parameter estimation using four sets of synthetic data, each with different numbers of data points, specifically $N=100, 200, 300,$ and $500$, taken from the smaller $q$.
    Figure \ref{fig:hc_fitting_result} shows the fitting results, with the synthetic data plots and MAP solutions.
    Figure \ref{fig:hc_posterior_result} presents the posteriors of the sampled parameters.
    In Fig. \ref{fig:hc_fitting_result}, when $N=100$, the data available from the wide-angle side are limited, leading to discrepancies between the estimated parameters and the theoretical curve.
    The posterior of the parameters $R_c$ and $dR_c$ when $N=100$ is bimodal, and is considered to be the estimation limit between $N=100$ and $200$. For all parameters, the larger the $q_{\rm{max}}$, the smaller the variance in the parameter posterior, indicating a tighter confidence interval for the estimations.
    Table \ref{tab:hc_estimated_values} shows the parameter estimation results  from the numerical experiments in this subsection.

    \begin{table}[hbtp]
        \caption{Parameter estimation results. The MAP solutions for the posterior of each estimated model parameter $\Theta_{\rm{MAP}}$ and ${\Theta_{\rm{MAP}}}^{+p}_{-m}$, where $p,m$ represents the differences between the upper and lower limits of the 99$\%$ confidence interval, are presented.}
        \label{tab:hc_estimated_values}
        \begin{tabular}{|l|c|c|c|c|c|}
            \hline
             & $R_c$ & $dR_c$ & $ b\,(\times 10^{-3})$ & $\rho_c\,(\times 10^{-2})$ & $\rho_s\,(\times 10^{-2})$ \\
            \hline
            \hline
            $N=500$   & $ 59.78^{+0.90}_{-0.83} $    & $ 10.27^{+0.89}_{-0.99} $   & $ 0.92^{+0.15}_{-0.14} $   & $ 100.13^{+0.29}_{-0.32} $     & $ 119.15^{+2.59}_{-1.88} $ \\
            $N=300$   & $ 60.07^{+0.98}_{-1.04} $     & $ 9.92^{+1.15}_{-1.10} $   & $ 0.74^{+0.34}_{-0.30} $   & $ 100.01^{+0.34}_{-0.33} $     & $ 120.23^{+3.37}_{-2.72} $ \\
            $N=200$   & $ 59.90^{+1.60}_{-1.41} $    & $ 10.14^{+1.55}_{-1.78} $   & $ 0.63^{+0.55}_{-0.42} $   & $ 100.10^{+0.36}_{-0.42} $     & $ 119.46^{+5.58}_{-3.24} $ \\
            $N=100$   & $ 59.24^{+7.99}_{-3.34} $   & $ 10.85^{+3.62}_{-10.43} $   & $ 0.45^{+1.85}_{-0.41} $   & $ 100.14^{+0.55}_{-4.81} $   & $ 118.05^{+937.49}_{-5.48} $ \\
            \hline
        \end{tabular}
    \end{table}

\newpage
\subsection{Assessing Estimation Accuracy with Reduced Data from the Low-Angle Side}
\begin{figure}[htbp]
        \centering
        \includegraphics[keepaspectratio, scale=0.47]{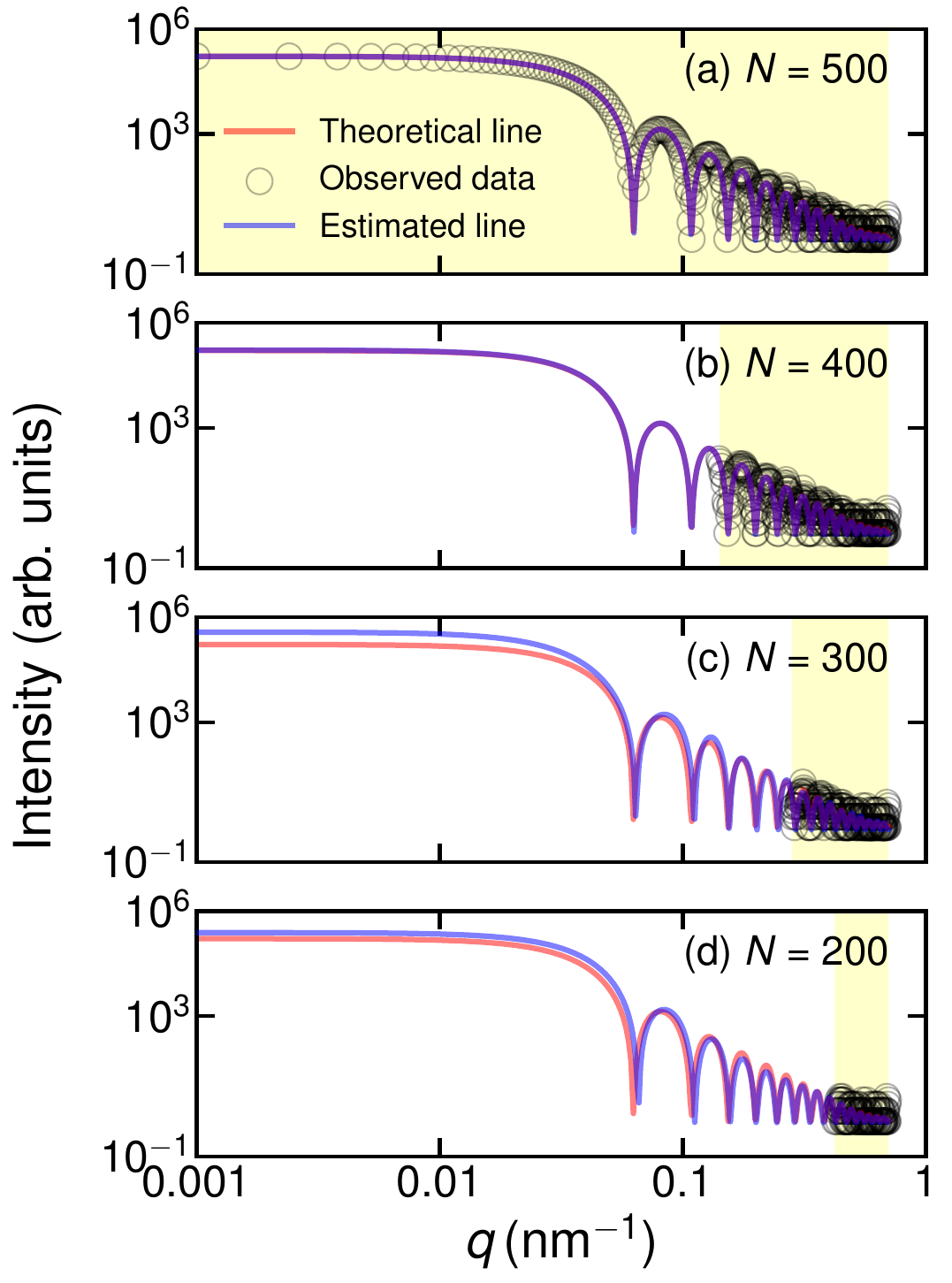}
        \caption{Plot of synthetic data at varying values of $q_{\rm{min}}$ and results of fitting with the MAP solution. The blue line indicates the estimated curve, the red line indicates the theoretical curve, and the open black circles denote the observed data. From top to bottom: $N=500, 400, 300,$ and $200$. The yellow-shaded area indicates the range over which data is available.}
        \label{fig:lc_fitting_result}
    \end{figure}

\begin{figure}[htbp]
        \includegraphics[keepaspectratio, scale=0.47]{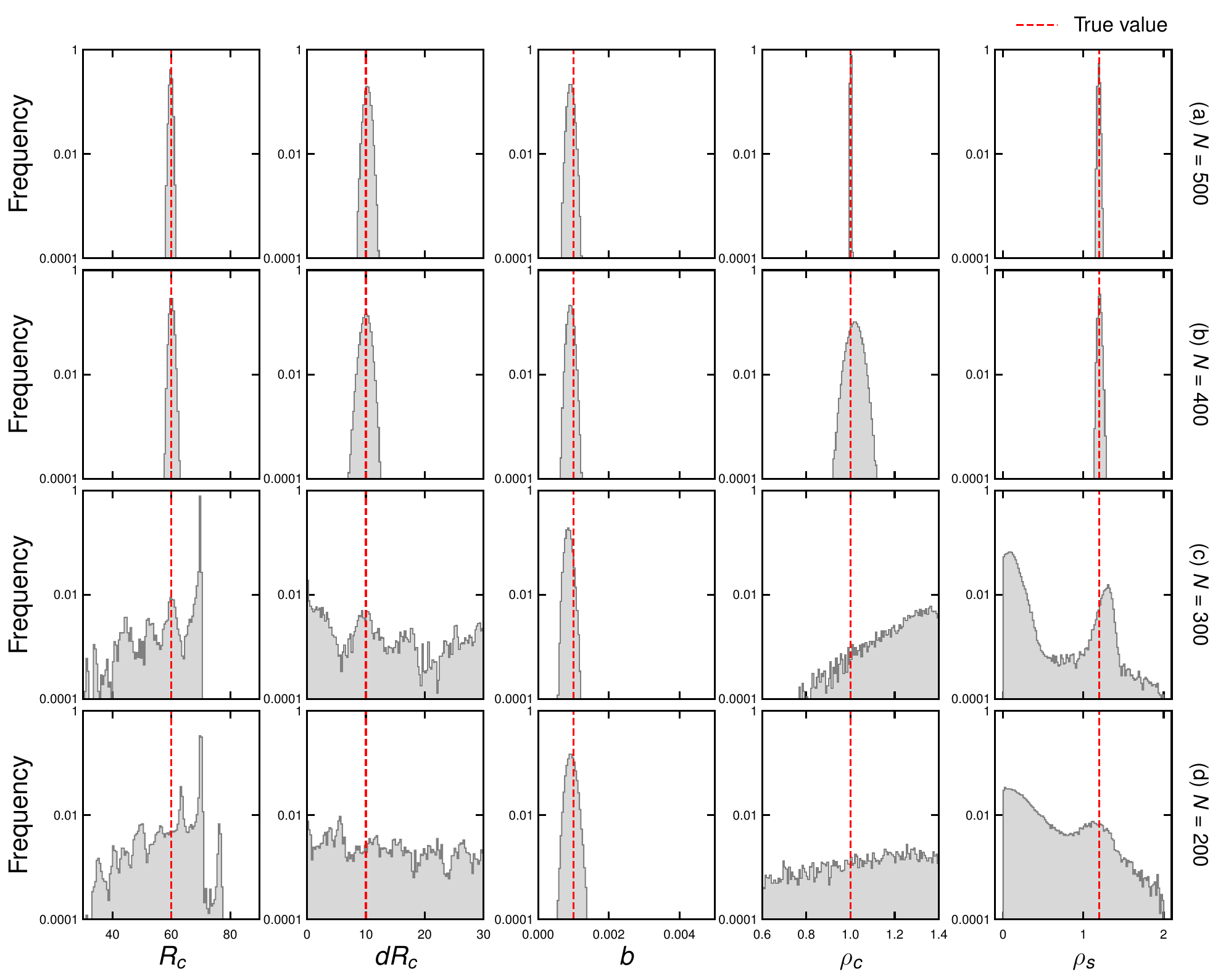}
        \caption{Parameters $R_c, dR_c, b, \rho_c$, and $\rho_s$ are arranged from left to right, with $t=1000$. The posteriors at sample sizes $N=500, 400, 300,$ and $200$ are shown from top to bottom, corresponding to the labels on the right. The resolution is maintained constant while $q_{\rm min}$ is varied. The dashed line indicates the true values. The vertical axis is logarithmic, and 120 bins are used for the histogram.}
        \label{fig:lc_posterior_result}
    \end{figure}

    In this subsection, we present the setup and results of a numerical experiment where data from the low-angle side are shaved off.
    First, data from the low-angle side were removed, and parameter estimation was performed using four sets of synthetic data with varying numbers of data points, specifically $N=200, 300, 400,$ and $500$, taken from the larger $q$.
    Figure \ref{fig:lc_fitting_result} shows the fitting results, including synthetic data plots and MAP solutions.
    Figure \ref{fig:lc_posterior_result} is the posterior of the sampled parameters.
    In Fig. \ref{fig:lc_fitting_result}, when $N=300$, there is little data on the low-angle side, and the parameter estimation on the low-angle side differs from the theoretical curve.
    The posterior of the parameters $\rho_s$ when $N=200, 300$ is bimodal, and is considered to be the estimation limit between $N=300$ and $400$. In the cases of $N=400$ and $500$, increasing the number of data points results in a smaller variance in the parameter posteriors, reflecting a reduced confidence interval of the estimations.
    Table \ref{tab:lc_estimated_values} shows the parameter estimation results from the numerical experiments in this subsection.


    \begin{table}[hbtp]
        \caption{Parameter estimation results. The MAP solutions for the posterior of each estimated model parameter $\Theta_{\rm{MAP}}$ and ${\Theta_{\rm{MAP}}}^{+p}_{-m}$, where $p,m$ represents the differences between the upper and lower limits of the 99$\%$ confidence interval, are presented.}
        \label{tab:lc_estimated_values}
        \begin{tabular}{|l|c|c|c|c|c|}
            \hline
             & $R_c$ & $dR_c$ & $ b\,(\times 10^{-3})$ & $\rho_c\,(\times 10^{-2})$ & $\rho_s\,(\times 10^{-2})$ \\
            \hline
            \hline
            $N=500$     & $ 59.78^{+0.84}_{-0.84} $      & $ 10.26^{+0.92}_{-0.92} $   & $ 0.92^{+0.15}_{-0.14} $        & $ 100.11^{+0.30}_{-0.29} $      & $ 119.23^{+2.32}_{-1.97} $ \\
            $N=400$     & $ 60.00^{+1.40}_{-1.12} $      & $ 10.05^{+1.21}_{-1.51} $   & $ 0.92^{+0.15}_{-0.15} $        & $ 101.78^{+5.44}_{-5.05} $      & $ 119.92^{+3.92}_{-2.98} $ \\
            $N=300$    & $ 69.79^{+0.21}_{-22.88} $   & $ 48.23^{+434.00}_{-48.17} $   & $ 0.83^{+0.20}_{-0.13} $   & $ 309.03^{+2648.53}_{-194.60} $    & $ 12.30^{+934.07}_{-11.53} $ \\
            $N=200$   & $ 59.74^{+10.83}_{-16.47} $    & $ 10.65^{+492.52}_{-9.33} $   & $ 0.93^{+0.23}_{-0.21} $     & $ 97.49^{+2784.02}_{-48.53} $   & $ 132.29^{+38.68}_{-130.82} $ \\
            \hline
        \end{tabular}
    \end{table}
\newpage

\section{Discussion}
In this study, we limited our numerical experiments to the specific condition $\rho_c > \rho_s > \rho_{\mathrm{solv}}$. However, this particular scenario represents only one of many possible configurations. Other conditions that warrant further investigation include:
\begin{itemize}
    \item $\rho_c < \rho_s < \rho_{\mathrm{solv}}$
    \item $\rho_c > \rho_s < \rho_{\mathrm{solv}}$
    \item $\rho_c > \rho_s > \rho_{\mathrm{solv}}$ (the condition examined in this study)
    \item $\rho_c < \rho_s > \rho_{\mathrm{solv}}$
    \item $\rho_c = \rho_s \neq \rho_{\mathrm{solv}}$
    \item $\rho_c \neq \rho_s = \rho_{\mathrm{solv}}$
    \item $\rho_c = \rho_{\mathrm{solv}} \neq \rho_s$
\end{itemize}
To firmly establish the superiority of Bayesian inference, it will be necessary to assess its performance under a broader array of such conditions. In future work, we plan to conduct more application-oriented numerical experiments and compare the parameter estimation results obtained from conventional methods with those derived from Bayesian inference.

Furthermore, determining whether a spherical or a core–shell model is more appropriate for a given dataset remains a crucial and challenging issue. Addressing this issue will require the development of a robust model selection framework, which we identify as a key avenue for future research.

\section{Conclusion}
\label{sec:conclusion}
In this study, we employed a Bayesian inference framework to investigate parameter estimation of the core–shell model in SAS experiments under various conditions. Using synthetic data generated from the theoretical expressions of the core–shell model, we demonstrated both scenarios in which parameter estimation is feasible and those in which estimation becomes challenging due to bimodal posterior distributions. As part of our future work, we plan to conduct more application-oriented numerical experiments under a wider range of experimental conditions and compare the results with those obtained using conventional approaches. Furthermore, selecting the appropriate model—whether a spherical or a core–shell model—is a significant and difficult issue in SAS analysis. To address the issue, we will also develop a model selection framework using synthetic datasets to determine which model is more suitable for given experimental data.

\bibliographystyle{unsrt}
\bibliography{reference}    


\end{document}